\documentclass[10pt]{article} 
\usepackage{amsfonts,amssymb,slashed,makeidx,latexsym,setspace}
\usepackage{graphicx,graphics,floatflt,amssymb,epsf,rotate,subfigure} 
\usepackage{times,cite,color,epsfig,epsf}
\usepackage{multirow}
\usepackage{hyperref}
\begin{document}

\begin{flushright}
SINP/TNP/2015/01, IPMU-15-0087
\end{flushright}

\begin{center}
{\Large \bf Focus Point Gauge Mediation with Incomplete Adjoint
  Messengers \\ \vspace{3pt} and Gauge Coupling Unification} \\
\vspace*{1cm} \renewcommand{\thefootnote}{\fnsymbol{footnote}} { {\sf
Gautam Bhattacharyya${}^{1}$}, {\sf Tsutomu T. Yanagida${}^{2}$}, and 
{\sf Norimi Yokozaki${}^{3}$}
}
\\
\vspace{10pt} {\small ${}^{1)}$ {\em Saha Institute of Nuclear
    Physics, 1/AF Bidhan Nagar, Kolkata 700064, India} \\ ${}^{2)}$
  {\em Kavli IPMU, TODIAS, University of Tokyo, Kashiwa 277-8583,
    Japan} \\${}^{3)}$ {\em INFN, Sezione di Roma, Piazzale A. Moro 2,
    I-00185 Roma, Italy} \\ } \normalsize
\end{center}

\begin{abstract}
As the mass limits on supersymmetric particles are gradually pushed to
higher values due to their continuing non-observation at the CERN LHC,
looking for focus point regions in the supersymmetric parameter space,
which shows considerably reduced fine-tuning, is increasingly more
important than ever. We explore this in the context of gauge mediated
supersymmetry breaking with messengers transforming in the adjoint
representation of the gauge group, namely, octet of color SU(3) and
triplet of weak SU(2). A distinctive feature of this scenario is that
the focus point is achieved by fixing a single combination of
parameters in the messenger sector, which is invariant under the
renormalization group evolution. Because of this invariance, the focus
point behavior is well under control once the relevant parameters are
fixed by a more fundamental theory. The observed Higgs boson mass is
explained with a relatively mild fine-tuning
$\Delta=60$\,-\,150. Interestingly, even in the presence of incomplete
messenger multiplets of the SU(5) GUT group, the gauge couplings still
unify perfectly, but at a scale which is one or two orders of
magnitude above the conventional GUT scale.  Because of this larger
unification scale, the colored Higgs multiplets become too heavy to
trigger proton decay at a rate larger than the experimentally allowed
limit.
\end{abstract}

\setcounter{footnote}{0}
\renewcommand{\thefootnote}{\arabic{footnote}}

\paragraph{\large Introduction:} Though still elusive, Supersymmetry
(SUSY), as a class of models, continues to be the leading candidate
for physics beyond the Standard Model (SM).  In addition to showing
the virtue of gauge coupling unification, supersymmetry provides a dynamical
origin of the {\em negative} mass-square of a neutral scalar that
triggers electroweak symmetry breaking (EWSB). As we know by now, the
origin of EWSB is completely explained if the scalar top (stop) mass
is around the weak scale.
However, the continuing absence of SUSY signals at the CERN Large
Hadron collider (LHC) has pushed up the gluino and squark masses to
larger than about 1.2-1.6 TeV~\cite{lhc_susy}.  Additionally, the
observed Higgs boson mass around 125 GeV~\cite{higgs_exp} in the SUSY
framework requires large radiative corrections from the
stops~\cite{higgs_rad}. This in turn necessitates the average stop
mass to be at least 3-5 TeV~\cite{higgs_3loop}, which is significantly
larger than the weak scale.  Consequently, settling the EWSB scale at
the correct value requires a large fine-tuning of the Higgs potential in
general.

Under these circumstances, the focus point SUSY~\cite{fp_org} (see
also \cite{fp_feng_recent} for recent discussions) deserves special
attention.  In this class of scenarios, one or more fixed ratios among
soft SUSY breaking masses are introduced, which lessens the
fine-tuning of the Higgs potential lending more credibility to the
natural explanation of the EWSB scale even if the SUSY particles turn
out to be very heavy.

Among focus point SUSY scenarios~\cite{fp_nonuniv,fp_gaugino,fp_e7,
  fp_e7_2}, the scenarios based on gauge mediation~\cite{gmsb}\,\footnote{
  For early attempts, see also Refs.~\cite{gmsb0}.
  } have
the advantage of suppressing the FCNC processes.\footnote{The focus
  point SUSY models based on gaugino mediation~\cite{fp_gaugino} and
  Higgs-gaugino mediation~\cite{fp_e7} also do not suffer from the
  SUSY FCNC problem. Moreover, the latter model can easily explain the
  muon $g-2$ anomaly~\cite{fp_e7_2}.}  In the context of gauge
mediation, the issue of focus point has been addressed in
Ref.~\cite{fp_gmsb}, where the numbers of the weakly and strongly
coupled messenger multiplets ($N_2$ and $N_3$, respectively) are
different from each other. Thanks to sizable cancellation between soft
mass parameters for particular choices of $N_2$ and $N_3$ during the
renormalization group running, the EWSB scale is realized with milder
fine-tuning. However, owing to the presence of these large number of
incomplete multiplets of the grand unified theory (GUT) group, the
gauge couplings do not unify.

The gauge coupling unification may be achieved non-trivially in a
framework where the messenger particles of gauge mediation transform
in the adjoint representation of the GUT group.  First, in
Ref.~\cite{yanagida-original}, it was shown that the presence of
adjoint matter multiplets with mass around $10^{13}$\,-\,$10^{14}$ GeV
can lead to gauge coupling unification around the string
scale~\cite{string}, which is one or two orders of magnitude above the
conventional GUT scale, {\em even if} the adjoint matters do not form
complete GUT multiplets. Subsequently, it was noticed that these
adjoint multiplets can be employed as messenger
superfields~\cite{amgmsb0} for gauge mediation that would generate
soft SUSY breaking masses.  Such adjoint gauge mediation scenarios
naturally lead to mass splitting among colored and uncolored particles
right at the messenger scale~\cite{amgmsb0,amgmsb1}.

We note at this point that in the context of SU(5) GUT, working with
adjoint messengers, namely, SU(3) octet and SU(2) triplet, which are
incomplete multiplets of SU(5), has a certain advantage over using
messengers of complete multiplets, e.g. SU(3) triplet and SU(2)
doublet.  In the latter case, the requirement of precise gauge
coupling unification demands that some colored Higgs multiplets weigh
around $10^{15}$-$10^{16}$ GeV~\cite{gut_p_decay}, which would prompt
unacceptably large proton decay. On the other hand, gauge coupling
unification with adjoint messengers would necessitate the colored
Higgs multiplets to hover around $10^{17}$-$10^{18}$ GeV due to the
larger unification scale, which is consistent with proton
lifetime~\cite{p_decay1, proton_exp}.

In this Letter, we exhibit how the fine-tuning of the EWSB can be
reduced by utilizing the mass splitting of the adjoint representation
messengers, more specifically, between the SU(3) octet and SU(2)
triplet messengers.  The focus point behavior is controlled by fixing
one single combination of the superpotential parameters.  Remarkably,
this combination is invariant under the renormalization group
evolution, i.e. it is stable against radiative corrections. Thus the
focus point behavior in adjoint messenger gauge mediation model is
more robust (assuming that the value of this combination is fixed by
some more fundamental physics) than other SUSY breaking scenarios in
the general class of minimal supersymmetric standard model (MSSM).  In
the latter scenarios, to reach the focus point region, various
relations among soft SUSY breaking and/or preserving (like $\mu$)
parameters need to be assumed which are neither invariant under
renormalization group evolution nor independent of the SUSY breaking
scale. This lends a substantial credibility to the attainment of focus
point in adjoint messenger gauge mediation models.

\paragraph{\large Adjoint messenger gauge mediation (AMGMSB):}
In the present scenario SUSY breaking is accomplished by gauge
mediation with messengers transforming in the adjoint representation
of the gauge group~\cite{amgmsb0,amgmsb1,fmyy}. These messengers
transform as $({\bf 8,1})$ and $({\bf 1,3)}$ under ${\rm SU(3)}_C \times
{\rm SU(2)}_L$ gauge group, and may have originated from the
non-Goldstone modes of the ${\bf 24}$ dimensional Higgs multiplet in
the SU(5) GUT gauge group. The resultant soft masses of weakly and
strongly interacting supersymmetric particles, which are significantly
different from those in minimal GMSB, allow for a significant reduction
of fine-tuning~\cite{fmyy}.  The superpotential in the messenger
sector is:
\begin{eqnarray}
\label{eq:mess1}
W_{\rm mess}= (M_8 + \lambda_8 Z ) {\rm Tr}(\Sigma_8^2) + (M_3 +
\lambda_3 Z ) {\rm Tr}(\Sigma_3^2) \, , 
\end{eqnarray}
where $Z$ is a spurion field whose $F$-term vacuum expectation value
(VEV) $F_Z$ breaks supersymmetry, whose effects are transmitted to the
observable sector via messenger loops.

Even though the messenger multiplets in our model are incomplete SU(5)
multiplets, the gauge coupling unification is still achieved for $M_3
\sim M_8 \sim 10^{13}$\,-\,$10^{14}$\,GeV
 at a scale somewhat higher
than the conventional $M_{\rm GUT} \simeq 10^{16}$
GeV~\cite{yanagida-original}, being around $M_{\rm str} \approx 5
\cdot 10^{17}$ GeV, which we call the string scale~\cite{string}. It
is expected that at this scale the gauge and gravitational couplings
are unified.

For illustration of gauge unification, we display the one-loop
beta-functions of the gauge couplings. The gauge couplings at $M_{\rm
  str}$ are given by
\begin{eqnarray}
\label{eq:rgemod}
\alpha_1^{-1} (M_{\rm str}) &=& \alpha_1^{-1} (m_{\rm SUSY}) -
\frac{b_1}{2\pi} \ln \frac{M_{\rm str}}{m_{\rm SUSY}}, \nonumber \\
\alpha_2^{-1} (M_{\rm str}) &=& \alpha_2^{-1} (m_{\rm SUSY})-
\frac{b_2+2}{2\pi} \ln \frac{M_{\rm str}}{m_{\rm SUSY}}  
+ \frac{2}{2\pi}\ln \frac{M_{3}}{m_{\rm SUSY}},  \\
\alpha_3^{-1} (M_{\rm str}) 
&=& \alpha_3^{-1} (m_{\rm SUSY}) 
- \frac{b_3+3}{2\pi} \ln \frac{M_{\rm str}}{m_{\rm SUSY}} 
+ \frac{3}{2\pi} \ln \frac{M_8}{m_{\rm SUSY}}, \nonumber
\end{eqnarray}
where $b_i=(33/5, 1, -3)$ is the coefficient of one-loop beta-function
for the gauge coupling $g_i$, and $m_{\rm SUSY}$ is the typical mass
scale of strongly interacting SUSY particles, defined here more
specifically as the stop mass scale $m_{\rm SUSY} \equiv (m_{{Q}_3}
m_{{\bar U}_3})^{1/2}$. It turns out that $\alpha_{1,2,3}^{-1} \simeq
(57, 31, 13)$ at $m_{\rm SUSY} = 3$ TeV.

From Eq.~(\ref{eq:rgemod}), we can write (following the discussion in
Ref.\cite{Hisano:1992mh})
\begin{eqnarray} 
(5\alpha_1^{-1} - 3\alpha_2^{-1} - 2\alpha_3^{-1})(m_{\rm SUSY}) &=&
  \frac{6}{\pi} \ln \left[\left(\frac{M_{\rm mess}}{m_{\rm
        SUSY}}\right)\left(\frac{M_{\rm str}}{m_{\rm
        SUSY}}\right)^2\right] \nonumber \\ 
(\alpha_1^{-1} -  3\alpha_2^{-1} +2\alpha_3^{-1})(m_{\rm SUSY}) &=& -\frac{6}{5\pi}
  \ln \left(\frac{M_{\rm str}}{m_{\rm SUSY}}\right) + \frac{3}{\pi}
  \ln \frac{M_{3}}{M_{8}} .\nonumber \\ \label{eq:gut}
\end{eqnarray} 
Using the above it is straightforward to obtain $M_{\rm str}^2 M_{\rm
  mess} \simeq M_{\rm GUT}^3$, where $M_{\rm mess} \equiv (M_3
M_8)^{1/2}$.  Requiring $M_{\rm str} \lesssim 10^{18}$ GeV, it follows
that $M_{\rm mess} \gtrsim 10^{12}$ GeV.  From the second equation of
Eq.~(\ref{eq:gut}), we see that the larger $M_{\rm str}$, or
equivalently smaller $M_{\rm mess}$, requires a larger ratio of
$M_3/M_8$ for the gauge coupling unification. For instance, for
$M_{\rm str}=10^{17}$ ($10^{18}$) GeV, one requires $M_3/M_8 \simeq 7
(18)$ at the one-loop level. We, however, employ two-loop
renormalization group equations (RGE) for the running of the gauge
couplings, which is displayed in Fig.~1.

\begin{floatingfigure}[r]{0.45\textwidth} 
\label{fig:gut}
\includegraphics[scale=1.0]{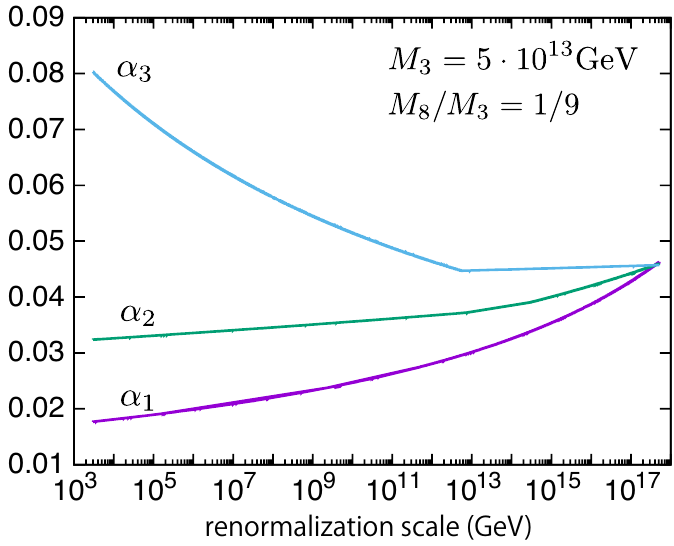}
\caption{Unification of the three gauge couplings at the two-loop
  level with ${\rm SU(3)}_C$ octet and ${\rm SU(2)}_L$ triplet
  messengers with their masses around $10^{13}$ GeV.  Here,
  $\alpha_s(M_Z) = 0.1185$ and $m_{\rm SUSY}=$ 3 TeV.}
\end{floatingfigure}

It is appropriate at this stage to highlight an advantage of using
adjoint messengers for gauge mediation in GUT framework, more
specifically, with SU(5) as the GUT group. The high scale spectra
invariably contain colored Higgs multiplets, namely, $H_C$ and $\bar
H_{ C}$, which belong to ${\bf 5}_H(= (H_u, H_C))$ and $\bar {\bf
  5}_H(= (H_d, \bar H_{C}))$ of SU(5), where $H_u$ $(H_d)$ denotes the
up-type (down-type) weak doublet Higgs multiplets.  The mass of the
colored Higgs multiplet $M_{H_C}$ is predicted to be around the
unification scale.\footnote{ The contributions of these colored Higgs
  states to gauge coupling evolution cannot be ignored if $M_{H_C}$ is
  smaller than $M_{\rm str}$. To account for their contributions, one
  must add $-(1/5\pi) \ln\left(M_{\rm str}/M_{H_c}\right)$ to
  $\alpha_1^{-1}$ and $-(1/2\pi) \ln\left(M_{\rm str}/M_{H_c}\right)$
  to $\alpha_3^{-1}$ in Eq.~(\ref{eq:rgemod})}  Adjoint messenger
gauge mediation has the distinct advantage of pushing the unification
scale beyond the conventional GUT scale to $M_{\rm str} (\sim M_{H_C})
= 10^{17}$-$10^{18}$ GeV, which can easily accommodate the
experimental constraints from the proton lifetime ~\cite{proton_exp}.
This is because the proton decay rate ($p \to K^+ \bar \nu$) is
suppressed by $1/M_{H_C}^2$~\cite{p_decay1}.

On the contrary, if the messengers are complete multiplets of SU(5),
the unification scale is $M_{\rm GUT} \sim10^{16}$ GeV (the
conventional scale), and then $M_{H_C} \sim M_{\rm GUT}$.  Moreover,
the precise gauge coupling unification requires $M_{H_C}$ to be
$10^{15}$-$10^{16}$ GeV. This necessitates inclusion of threshold
corrections to the gauge couplings, namely, $-(1/5\pi) \ln\left(M_{\rm
  GUT}/M_{H_c}\right)$ to $\alpha_1^{-1}$ and $-(1/2\pi)
\ln\left(M_{\rm GUT}/M_{H_c}\right)$ to $\alpha_3^{-1}$.  Then the
proton decay rate would overshoot the experimental limit for (sub-)TeV
scale SUSY~\cite{gut_p_decay}.

With these messenger multiplets, the gaugino masses from the messenger
loops at the scale $M_{\rm mess}$ are 
    \begin{eqnarray}
M_{\tilde B} \simeq 0 \, , ~~
M_{\tilde W} \simeq \frac{g_2^2}{16\pi^2} (2 \Lambda_3) \, ,~~
M_{\tilde g} \simeq  \frac{g_3^2}{16\pi^2} (3 \Lambda_8) \, , \label{eq:mass1}
\end{eqnarray}
where $\Lambda_3 \equiv \lambda_3 \left<F_Z\right>/M_3$ and $\Lambda_8
\equiv \lambda_8 \left<F_Z\right>/M_8$, provided that $\lambda_{3}
\left<Z\right>$ and $\lambda_8 \left<Z\right>$ are much smaller than
$M_3$ and $M_8$, respectively. The sfermion masses at $M_{\rm mess}$
are given by
  \begin{eqnarray}
m_{{Q}}^2 &\simeq& \frac{2}{(16\pi^2)^2} \left[\frac{4}{3} g_3^4 (3
  \Lambda_8^2) + \frac{3}{4} g_2^4 (2 \Lambda_3^2) \right] \, , ~~
m_{\bar D}^2 = m_{\tilde{U}}^2 \simeq \frac{2}{(16\pi^2)^2}
\frac{4}{3} g_3^4 (3 \Lambda_8^2), \nonumber \\ 
m_{{L}}^2 &=& m_{H_u}^2 = m_{H_d}^2 \simeq
\frac{2}{(16\pi^2)^2} \frac{3}{4} g_2^4 (2 \Lambda_3^2) \, , ~~
m_{\bar E}^2 \simeq 0. \label{eq:mass2}
\end{eqnarray}

\vspace{7pt}
One can see that the bino and right-handed sleptons are massless,
since there is no messenger field charged under the $U(1)_Y$ gauge
group.  In order to give masses to the right-handed sleptons, we
consider the minimal Kahler for the MSSM matter multiplets and the
spurion $Z$.  Then the MSSM matter fields receive a common mass $m_0$
from the supergravity scalar potential, which is equal to the
gravitino mass $m_{3/2} = \left<F_Z\right>/(\sqrt{3} M_P)$.

In this setup, the gluino mass at the soft SUSY breaking mass scale
($\sim$ TeV) is 
\begin{eqnarray}
M_{\tilde g}(m_{\rm SUSY}) = \frac{\alpha_3(m_{\rm SUSY})}{4\pi} (3
\Lambda_8) ~\simeq~ 4.0 \,{\rm TeV} \cdot
\left(\frac{\lambda_8}{0.001}\right) \left(\frac{m_{3/2}}{500\,{\rm
    GeV}}\right) \left(\frac{M_{8}}{10^{13}\,{\rm GeV}}\right)^{-1}.
\end{eqnarray}

The bino can get a mass from the gauge kinetic function:
\begin{eqnarray}
\mathcal{L}\ni \frac{1}{4 g_1^2} \int d^2\theta \left(1 - \frac{ 2 k
  Z}{M_P}\right) W_{\alpha}^1 W^{\alpha\, 1} + {\rm h.c.}
\end{eqnarray}
Then, 
\begin{eqnarray}
M_{\tilde B}(M_{\rm str})= \frac{k \left<F_Z\right>}{M_P} = \sqrt{3} k m_{3/2}.
\end{eqnarray}

Alternatively, one can consider the sequestered form of the Kahler
potential, which ensures the absence of FCNC.  In this case, the
right-handed slepton masses are generated by the bino-loop, which is
nothing but the gaugino mediation mechanism. Another option is to
introduce a pair of ${\bf 5}$ and $\bar {\bf 5}$ messengers to
generate the bino and right handed slepton masses, which would also
contribute to other masses.  In this Letter, for simplicity, we work
with only SU(3) octet and SU(2) triplet adjoint messengers and stick
to the case of the minimal Kahler potential, as mentioned above.

\paragraph{\large Focus point in the AMGMSB:}
We consider the fine-tuning of the EWSB scale with soft masses
generated from these adjoint messengers.  The EWSB conditions are
given by
\begin{eqnarray}
\frac{g_1^2 + g_2^2}{4} v^2 &=& \Bigl[ -\mu^2 - \frac{(m_{H_u}^2 +
    \frac{1}{2 v_u}\frac{\partial \Delta V}{\partial v_u} )
    \tan^2\beta}{\tan^2\beta-1} + \, \frac{m_{H_d}^2 + \frac{1}{2
      v_d}\frac{\partial \Delta V}{\partial v_d} }{\tan^2\beta-1}
  \Bigr]_{m_{\rm SUSY}}, \nonumber \\ \frac{\tan^2\beta+1}{\tan\beta}
&=& \Bigl[\frac{1}{B \mu}\Bigl( m_{H_u}^2 +\frac{1}{2
    v_u}\frac{\partial \Delta V}{\partial v_u} + m_{H_d}^2 +
  \frac{1}{2 v_d}\frac{\partial \Delta V}{\partial v_d} + 2\mu^2
  \Bigr) \Bigr]_{m_{\rm SUSY}}, \label{eq:ewsb}
\end{eqnarray}
where $\Delta V$ denotes an one-loop correction to the Higgs
potential, and $m_{H_u}^2$ and $m_{H_d}^2$ are the soft masses for the
up-type and down-type Higgs, respectively.  The Higgsino mass
parameter is denoted by $\mu$, and $B \mu$ is a soft SUSY breaking
parameter of the Higgs bi-linear term.  The above equations tell us
that the EWSB scale is determined dominantly by $\mu^2$ and
$[m_{H_u}^2 + 1/(2 v_u)({\partial \Delta V}/{\partial v_u})]$ for
large $\tan\beta$\,($\equiv \left<H_u^0\right>/\left<H_d^0\right>$).

Now we consider renormalization group running of $m_{H_u}^2$ from the high scale to weak scale.
To understand the behavior of this running intuitively, 
we first demonstrate it  using approximate analytic solutions of RGEs.
The Higgs soft masses receive negative contributions from stop and gluino loops and positive contributions from the wino loop. 
The dominant negative contributions induced by the top-Yukawa coupling $y_t$ are~\footnote{
The difference of the coefficients in front of $m_{Q}^2$ and $m_{\bar U}^2$ between Eq.\,(\ref{eq:neg_cont}) and Eq.\,(\ref{mhu3tev}) arises from ${\rm U(1)}_Y$ contributions:
\begin{eqnarray}
({16\pi^2})\frac{d m_{H_u}^2}{d t} \ni \frac{3}{5}g_1^2 [{\rm Tr}(m_{Q}^2 - 2 m_{\bar U}^2)  + \dots].
\end{eqnarray}

However, the above ${\rm U(1)}_Y$ contributions are eventually canceled out in the most of the gauge mediation models when their effects on each individual soft masses are summed up.
}
\begin{eqnarray}
(m_{H_u}^{2} (Q_r) )_{\rm neg}  \simeq \frac{k-1}{2} \left[ m_{Q_3}^2 (M_{\rm mess}) + m_{\bar U_3}^2(M_{\rm mess}) \right] - k_{\tilde g} M_{\tilde g}^2 (M_{\rm mess}), \label{eq:neg_cont}
\end{eqnarray}
where $Q_r$ is the renormalization scale taken to be the stop mass scale and
\begin{eqnarray}
k &=& \exp\left[ \int_0^t  \left( \frac{3 y_t^2(t') }{4\pi^2} \right) dt' \right], \ \ {\rm with} \ \ t = \ln ({Q_r}/M_{\rm mess}), \nonumber \\
k_{\tilde g} &=& \int_0^t dt' \, \frac{y_t^2(t') g_3^2 (M_{\rm mess})}{2\pi^4} t' \left[\frac{1-\eta_3 t' /2}{(1-\eta_3 t')^2}\right]
\nonumber \\
&-& \int_0^t dt' \, \frac{y_t^2(t') g_3^4 (M_{\rm mess})}{6\pi^6} \frac{t'^2}{(1-\eta_3 t')^2},
\ \ {\rm with} \ \ \eta_3 = -3\frac{g_3^2(M_{\rm mess})}{8\pi^2} .
\end{eqnarray}
In addition to the above negative contributions, there are positive contributions arising from the wino loop and tree-level Higgs soft mass:
\begin{eqnarray}
(m_{H_u}^{2} (Q_r) )_{\rm pos}  \simeq \frac{k+1}{2} m_{H_u}^2 (M_{\rm mess}) +  k_{\tilde W} M_{\tilde W}^2(M_{\rm mess}),
\end{eqnarray}
where we show the only dominant contributions and 
\begin{eqnarray}
k_{\tilde W} &=& -\frac{3 g_2^2 (M_{\rm mess})}{8\pi^2} t  \, \left[ \frac{1-\eta_2 t /2}{(1-\eta_2 t)^2}\right] \nonumber \\
&-& \int_0^t dt' \, \frac{9y_t^2(t') g_2^2 (M_{\rm mess})}{32\pi^4 } \,  t' \left[ \frac{1-\eta_2 t'/2}{(1-\eta_2 t')^2} \right]
\ , 
 \ \ {\rm with} \ \ \eta_2 = \frac{g_2^2(M_{\rm mess})}{8\pi^2 } .
\end{eqnarray}
The sizes of the coefficients are $k\sim 0.4$, $k_{\tilde g} \sim 0.7$ and $k_{\tilde W} \sim 0.2$. Therefore in the case of the minimal GMSB with ${\bf 5}$ and $\bar {\bf 5}$ messengers, the negative contributions substantially dominate over the positive contributions, leading to only a small cancellation.
One thus needs larger $M_{\tilde W} (M_{\rm mess})$ and/or $m_{H_u}^2 (M_{\rm mess})$ to obtain a sizable cancellation leading to small $m_{H_u}^2$ at the soft mass scale. We will see below how it is achieved in our scenario.

 Now we evaluate the value of $m_{H_u}^2$ at $m_{\rm SUSY}$ more precisely by numerically solving
 two-loop RGEs ~\cite{twoloop_rge}.  By
taking $M_{\rm mess}=10^{13}$ GeV, $\tan\beta=15$, $m_{t}({\rm
  pole})=173.34$ GeV and $\alpha_s(m_Z)=0.1185$, we obtain 
\begin{eqnarray}
m_{H_u}^2(3 {\rm TeV}) &=& 0.704 m_{H_u}^2 + 0.019 m_{H_d}^2 \nonumber
\\ &-& 0.336 m_{Q}^2 -0.167 m_{\bar U}^2 -0.056 m_{\bar E}^2 \nonumber
\\ &+& 0.055 m_{L}^2 - 0.054m_{\bar D}^2 \nonumber \\ &+& 0.011
M_{\tilde B}^2 + 0.192 M_{\tilde W}^2 -0.727 M_{\tilde g}^2 \nonumber
\\ &-& 0.003 M_{\tilde B} M_{\tilde W} -0.062 M_{\tilde W}M_{\tilde g}
- 0.010 M_{\tilde B} M_{\tilde g},
\label{mhu3tev}
\end{eqnarray}
where soft SUSY
breaking mass parameters in the right hand side of Eq.~(\ref{mhu3tev})
are defined at $M_{\rm mess}$.  By using Eqs.(\ref{eq:mass1}) and
(\ref{eq:mass2}), we obtain ($r_3 \equiv \Lambda_3/\Lambda_8$)
\begin{eqnarray}
m_{H_u}^2(3 {\rm TeV}) \simeq [0.165 \, r_3^2 -0.035 \, r_3 - 1.222]
M_{\tilde g}^2 \, .  
\end{eqnarray}
Note that $m_{H_u}^2(3\,{\rm TeV})$ nearly vanishes for $r_3 \simeq
2.8, -2.6$, i.e. we reach a focus point region.  Here, we have
neglected the contribution from the universal scalar mass. In fact,
this contribution is rather small as $m_{H_u}^2(3\,{\rm TeV}) \ni
0.164 m_{0}^2$.  Here we make a crucial observation that the ratio
$r_3 \simeq \lambda_3 M_8/(\lambda_8 M_3)$ is RGE invariant:
\begin{eqnarray}
\lambda_{(3,8)}(t) &=& \lambda_{(3,8)}(t_0) \exp \left[ \int_{t_0}^{t}
  dt' (\gamma_Z + 2 \gamma_{\Sigma_{(3,8)}}) \right] , \nonumber
\\ M_{(3,8)}(t) &=& M_{(3,8)}(t_0) \exp \left[ \int_{t_0}^{t} dt' (2
  \gamma_{\Sigma_{(3,8)}}) \right],
\end{eqnarray}
where $\gamma_i$ is the anomalous dimension of the field $i$. It
immediately follows that 
\begin{eqnarray}
\frac{\lambda_3(t) M_8(t)}{\lambda_8(t) M_3(t)} = \frac{\lambda_3(t_0)
  M_8(t_0)}{\lambda_8(t_0) M_3(t_0)}.
\end{eqnarray}
Once the ratio $r_3$ is fixed by some fundamental physics, it is
stable against radiative corrections, i.e. invariant under RGE
running. This unique property lends significant reliability and
robustness to our scenario over other competitive focus point SUSY
models.

\begin{figure}[!t]
\label{fig:delta}
\begin{center}
\includegraphics[scale=1.05]{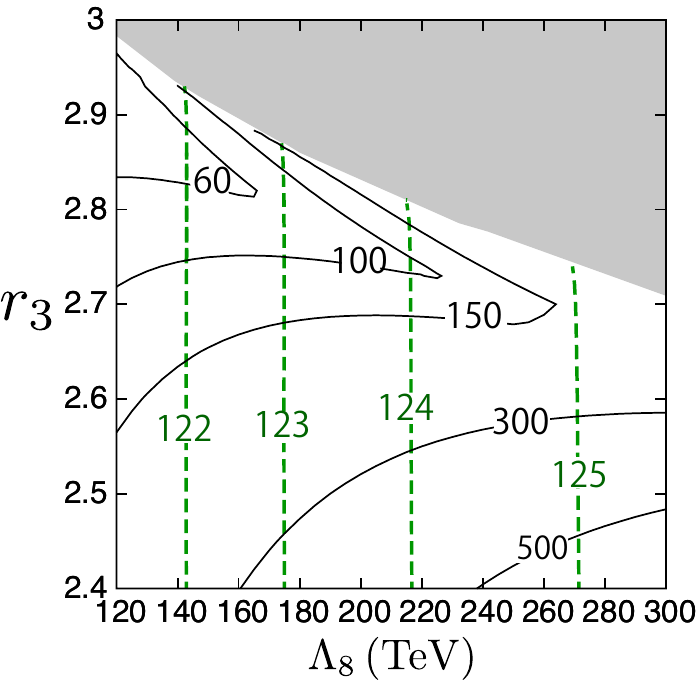}
\includegraphics[scale=1.05]{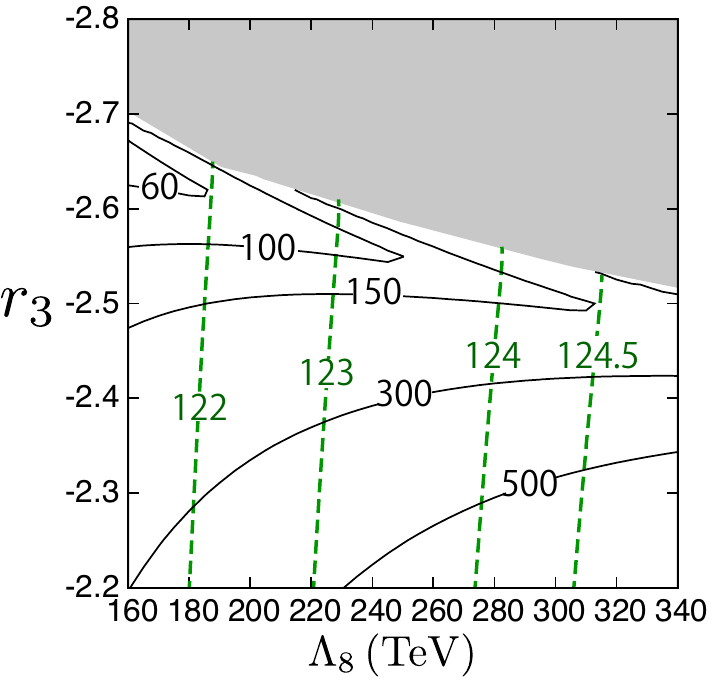}
\caption{Contours of $\Delta$ (solid) and $m_{h}/{\rm GeV}$
  (dashed). In the gray region, the EWSB does not occur.  
  The messenger scale is taken as $M_{\rm mess}=10^{13}$ GeV.
  Here,
  $\tan\beta=15$, $m_t({\rm pole})=173.34$\,GeV and $\alpha_s(M_Z) =
  0.1185$. }
\end{center}
\end{figure}

\begin{figure}[!t]
\label{fig:delta_mgmsb}
\begin{center}
\includegraphics[scale=1.05]{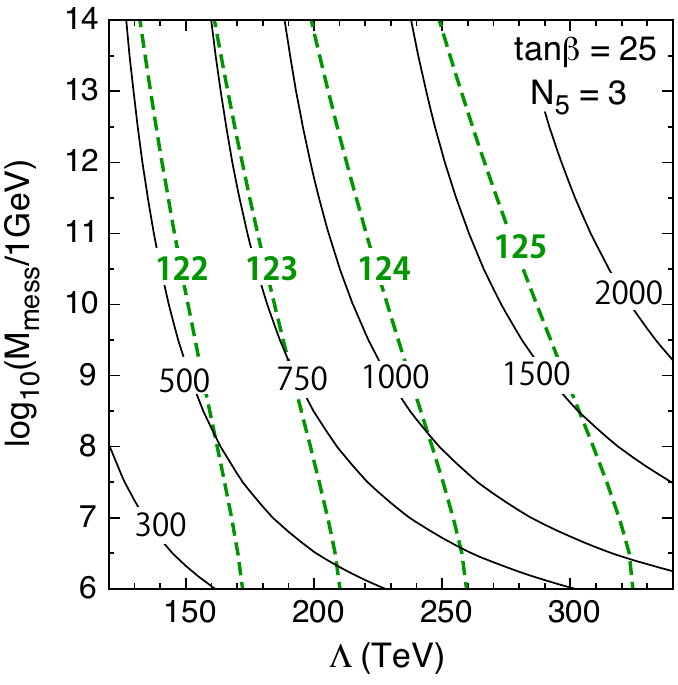}
\caption{Contours of $\Delta$ (solid) and $m_{h}/{\rm GeV}$ (dashed)
  in minimal GMSB (i.e. with ${\bf 5}$ and ${\bf \bar 5}$ messengers
  of SU(5) GUT).  We set $\tan\beta=25$ and $N_5=3$. Other parameters
  are same as in Fig. 2.}
\end{center}
\end{figure}

\begin{table}[t]
    \caption{Sample mass spectra. We take $M_{\rm mess}=10^{13}$ GeV. }
      \begin{center}
    \begin{tabular}{  c | c  }
    {\bf P1} & \\
    \hline
    $\Lambda_8$ & 180 TeV \\
    $r_3$ & 2.8 \\
    $\tan\beta$ & 15 \\
    $M_{1}(M_{\rm mess})$ & 250 GeV \\
    $m_0(M_{\rm mess})$ & 450 GeV \\
    \hline
\hline    
    $m_{h}$ & 123.1\,GeV \\
    $\Delta $ & 69 \\
    $|\Delta_{\mu}| $ & 61 \\
    $\mu $ & 538 GeV \\
\hline
    $m_{\rm gluino}$ & 3.6 TeV \\
    $m_{\rm squark}  $ & 3.4 - 4.5 TeV \\
    $m_{\rm stop}  $ & 2.2, \, 4.1 TeV \\
    $m_{\tilde{e}_L} (m_{\tilde{\mu}_L})$ & 3.1 TeV\\
    $m_{\tilde{e}_R} (m_{\tilde{\mu}_R})$ & 473 GeV \\
    $m_{\tilde{\tau}_1}$ & 221 GeV \\
    $m_{\chi_1^0}$ & 128 GeV \\
     $m_{\chi_1^{\pm}}$ & 550 GeV \\
     $m_{\chi_2^{\pm}}$ & 2.6 TeV \\
    \end{tabular}
        \begin{tabular}{  c | c  }
     {\bf P2} & \\
    \hline
    $\Lambda_8$ & 280 TeV \\
    $r_3$ & 8/3 \\
    $\tan\beta$ & 15 \\
    $M_{1}(M_{\rm mess})$ & 250 GeV \\
    $m_0(M_{\rm mess})$ & 700 GeV \\
    \hline
\hline    
    $m_{h}$ & 125.1\,GeV \\
    $\Delta $ & 156 \\
    $|\Delta_{\mu}| $ & 156 \\
     $\mu $ & 850 GeV \\
\hline
    $m_{\rm gluino}$ & 5.4 TeV \\
    $m_{\rm squark}  $ & 5.1 - 6.7 TeV \\
    $m_{\rm stop}  $ & 3.4, \, 6.2 TeV \\
    $m_{\tilde{e}_L} (m_{\tilde{\mu}_L})$ & 4.5 TeV\\
    $m_{\tilde{e}_R} (m_{\tilde{\mu}_R})$ & 727 GeV \\
    $m_{\tilde{\tau}_1}$ & 399 GeV \\
    $m_{\chi_1^0}$ & 124 GeV \\
     $m_{\chi_1^{\pm}}$ & 870 GeV \\
     $m_{\chi_2^{\pm}}$ & 3.8 TeV \\
    \end{tabular}
        \begin{tabular}{  c | c  }
     {\bf P3} & \\
    \hline
    $\Lambda_8$ & 230 TeV \\
    $r_3$ & -2.55 \\
    $\tan\beta$ & 15 \\
    $M_{1}(M_{\rm mess})$ & 250 GeV \\
    $m_0(M_{\rm mess})$ & 600 GeV \\
    \hline
\hline    
    $m_{h}$ & 123.0\,GeV \\
    $\Delta $ & 91 \\
    $|\Delta_{\mu}| $ & 91 \\
     $\mu $ & 652 GeV \\
\hline
    $m_{\rm gluino}$ & 4.5 TeV \\
    $m_{\rm squark}  $ & 4.2 - 5.5 TeV \\
    $m_{\rm stop}  $ & 3.1, \, 5.1 TeV \\
    $m_{\tilde{e}_L} (m_{\tilde{\mu}_L})$ & 3.6 TeV\\
    $m_{\tilde{e}_R} (m_{\tilde{\mu}_R})$ & 618 GeV \\
    $m_{\tilde{\tau}_1}$ & 394 GeV \\
    $m_{\chi_1^0}$ & 131 GeV \\
     $m_{\chi_1^{\pm}}$ & 670 GeV \\
      $m_{\chi_2^{\pm}}$ & 3.1 TeV \\
    \end{tabular}
  \label{table:mass}
  \end{center}
\end{table}

Now, we estimate the fine-tuning of the EWSB scale using the following
measure~\cite{ft_measure}:
\begin{eqnarray}
\Delta= {\rm max} \, \{ |\Delta_a| \}, \ \ \Delta_a =
\left[\frac{\partial \ln v}{\partial \ln |F_Z|}, \frac{\partial \ln
    v}{\partial \ln \mu}, \frac{\partial \ln v}{\partial \ln B_0},
  \frac{\partial \ln v}{\partial \ln M_1}, \frac{\partial \ln
    v}{\partial \ln m_0} \right]_{v= v_{\rm obs}},
\end{eqnarray}
where $v_{\rm obs} \simeq 174.1$ GeV and $B_0$ is the scalar potential
$B$-term at the messenger scale, which may, for example, be generated
by the Giudice-Masiero mechanism~\cite{mu_bmu} or from a constant term
in the superpotential. 

In Fig.~2, we show the contours of $\Delta$ and $m_{h}$. The Higgs
boson mass is calculated using {\tt FeynHiggs
  2.10.3}~\cite{feynhiggs}, and $\Delta$ is evaluated utilizing {\tt
  SOFTSUSY 3.6.1}~\cite{softsusy}.  To avoid the tachyonic stau, we
take the universal scalar masses at $M_{\rm mess}$ as~\footnote{
  Strictly speaking, the universal scalar masses should be taken at
  $M_{\rm str}$. However, it makes only a small difference.}
\begin{eqnarray}
m_{0}(M_{\rm mess}) = \left(\frac{\Lambda_8}{180\, {\rm TeV}}\right)
500 {\rm GeV}.
\end{eqnarray}
Also, the bino mass is regarded as an input parameter at $M_{\rm
  mess}$, and taken as $M_{\tilde B}(M_{\rm mess})=250$ GeV. The sign
of the $\mu$-parameter is taken to be positive. In the gray region
with large $|r_3|$, the EWSB does not occur. One can see the observed
Higgs boson mass is explained with $\Delta=60-150$ for $r_3 \sim 2.8$
(Fig.~2, left panel).  When $r_3$ is negative (Fig.~2, right-panel),
the required fine-tuning to reach the correct Higgs boson mass is
slightly larger than the positive $r_3$ case.
These results can be compared to the minimal GMSB case (with only
${\bf 5}$ and ${\bf \bar 5}$ messengers), shown in Fig.~3.  Demanding
$m_{h}>123$ GeV, the required $\Delta$ is around $750$-$1500$ for
$M_{\rm mess} \gtrsim 10^9$ GeV. For this plot we have taken the
number of ${\bf 5}$ and ${\bf \bar 5}$ pairs to be $N_5 = 3$, though
the required $\Delta$ does not significantly depend on this choice.  Comparing
Fig.~3 with Fig.~2 it is clear that our adjoint messenger model in the
focus point region for $r_3 \sim 2.8$ is significantly less tuned than
minimal GMSB.

Finally we show some sample spectrum in Table.~\ref{table:mass}.  One
can see that the stau can be light as 200-400 GeV, which may be
testable at the LHC depending on the bino and Higgsino masses.
Admittedly, the bino-like lightest neutralino may give rise to too
large relic density causing over-closure of the universe.  This can be
avoided by tuning on a tiny amount of $R$-parity violation.  In this
case, axion could become a potential dark matter candidate.

\paragraph{\large Conclusions:}

We have considered a gauge mediated SUSY breaking scenario with
messengers transforming in the adjoint representation of the gauge
group as color octet and weak triplet.  We have shown that focus point
exists in this framework. The fine-tuning of the EWSB scale is
considerably reduced in the focus point region: $\Delta=60$\,-\,$150$,
while explaining the observed Higgs boson mass around 125 GeV. In
fact, the fine-tuning is considerably reduced in our scenario compared
to that in minimal gauge mediation.  Two distinctive features
attribute a substantial credibility to our scenario: ($i$) a single
combination of messenger sector parameters, which is RGE invariant,
controls the focus point. This means that the focus point behavior is
stable once a more fundamental theory fixes that combination; ($ii$)
the special feature of color octet and weak triplet adjoint messengers
triggering late gauge unification renders consistency of the scenario
with colored Higgs mediated proton decay constraints.

\noindent {\bf Acknowledgements:}~ We thank Hajime Fukuda and Hitoshi
Murayama for useful discussions.  G.B. thanks Kavli IPMU for
hospitality when the work was done.  This work is supported by
Grants-in-Aid for Scientific Research from the Ministry of Education,
Culture, Sports, Science, and Technology (MEXT), Japan, No. 26104009,
Grant-in-Aid No. 26287039 from the Japan Society for the Promotion of
Science (JSPS), and the World Premier International Research Center
Initiative (WPI), MEXT, Japan (T. T. Y.).  The research leading to
these results has received funding from the European Research Council
under the European Unions Seventh Framework Programme (FP/2007-2013) /
ERC Grant Agreement n. 279972 “NPFlavour” (N. Y.).


\end{document}